\documentclass[twocolumn,showpacs,aps,prd,superscriptaddress,nofootinbib]{revtex4}

\usepackage{graphicx}

\usepackage{dcolumn}

\usepackage{epsfig}

\RequirePackage{xspace}

\usepackage{relsize}
\def\babar{\mbox{\slshape B\kern-0.1em{\smaller A}\kern-0.1em
    B\kern-0.1em{\smaller A\kern-0.2em R}}}

\def\epem       {\ensuremath{e^+e^-}\xspace}

\def\ubar  {\ensuremath{\overline u}\xspace}

\def\cbar  {\ensuremath{\overline c}\xspace}

\def\Kbar  {\kern 0.2em\overline{\kern -0.2em K}{}\xspace}

\def\Kz    {\ensuremath{K^0}\xspace}
\def\Kzb   {\ensuremath{\Kbar^0}\xspace}
\def\KzKzb {\ensuremath{\Kz \kern -0.16em \Kzb}\xspace}
\def\Kp    {\ensuremath{K^+}\xspace}
\def\Km    {\ensuremath{K^-}\xspace}

\def\KpKm  {\ensuremath{\Kp \kern -0.16em \Km}\xspace}
\def\KS    {\ensuremath{K^0_{\scriptscriptstyle S}}\xspace}

\def\Dbar    {\kern 0.2em\overline{\kern -0.2em D}{}\xspace}

\def\Dz      {\ensuremath{D^0}\xspace}
\def\Dzb     {\ensuremath{\Dbar^0}\xspace}
\def\DzDzb   {\ensuremath{\Dz {\kern -0.16em \Dzb}}\xspace}
\def\Dp      {\ensuremath{D^+}\xspace}
\def\Dm      {\ensuremath{D^-}\xspace}

\def\DpDm    {\ensuremath{\Dp {\kern -0.16em \Dm}}\xspace}

\def\Bbar    {\kern 0.18em\overline{\kern -0.18em B}{}\xspace}

\def\BB      {\ensuremath{B\Bbar}\xspace} 
\def\Bz      {\ensuremath{B^0}\xspace}
\def\Bzb     {\ensuremath{\Bbar^0}\xspace}
\def\BzBzb   {\ensuremath{\Bz {\kern -0.16em \Bzb}}\xspace}
\def\Bu      {\ensuremath{B^+}\xspace}
\def\Bub     {\ensuremath{B^-}\xspace}

\def\BpBm    {\ensuremath{\Bu {\kern -0.16em \Bub}}\xspace}
\def\Bs      {\ensuremath{B_s}\xspace}
\def\Bsb     {\ensuremath{\Bbar_s}\xspace}

\def\BorBbar    {\kern 0.18em\optbar{\kern -0.18em B}{}\xspace}
\def\DorDbar    {\kern 0.18em\optbar{\kern -0.18em D}{}\xspace}
\def\KorKbar    {\kern 0.18em\optbar{\kern -0.18em K}{}\xspace}

\mathchardef\Upsilon="7107
\def\Y#1S{\ensuremath{\Upsilon{(#1S)}}\xspace}

\mathchardef\Deltares="7101
\mathchardef\Xi="7104
\mathchardef\Lambda="7103
\mathchardef\Sigma="7106
\mathchardef\Omega="710A

\def\Deltabar{\kern 0.25em\overline{\kern -0.25em \Deltares}{}\xspace}
\def\Lbar{\kern 0.2em\overline{\kern -0.2em\Lambda\kern 0.05em}\kern-0.05em{}\xspace}
\def\Sigbar{\kern 0.2em\overline{\kern -0.2em \Sigma}{}\xspace}
\def\Xibar{\kern 0.2em\overline{\kern -0.2em \Xi}{}\xspace}
\def\Obar{\kern 0.2em\overline{\kern -0.2em \Omega}{}\xspace}
\def\Nbar{\kern 0.2em\overline{\kern -0.2em N}{}\xspace}
\def\Xb{\kern 0.2em\overline{\kern -0.2em X}{}\xspace}

\def\BR         {{\ensuremath{\cal B}\xspace}}

\newcommand{\tev}{\ensuremath{\mathrm{\,Te\kern -0.1em V}}\xspace}
\newcommand{\gev}{\ensuremath{\mathrm{\,Ge\kern -0.1em V}}\xspace}
\newcommand{\mev}{\ensuremath{\mathrm{\,Me\kern -0.1em V}}\xspace}
\newcommand{\kev}{\ensuremath{\mathrm{\,ke\kern -0.1em V}}\xspace}
\newcommand{\ev}{\ensuremath{\mathrm{\,e\kern -0.1em V}}\xspace}
\newcommand{\gevc}{\ensuremath{{\mathrm{\,Ge\kern -0.1em V\!/}c}}\xspace}
\newcommand{\mevc}{\ensuremath{{\mathrm{\,Me\kern -0.1em V\!/}c}}\xspace}
\newcommand{\gevcc}{\ensuremath{{\mathrm{\,Ge\kern -0.1em V\!/}c^2}}\xspace}
\newcommand{\mevcc}{\ensuremath{{\mathrm{\,Me\kern -0.1em V\!/}c^2}}\xspace}

\def\mus  {\ensuremath{\rm \,\mus}\xspace}

\def\fs   {\ensuremath{\rm \,fs}\xspace}

\def\mus        {\ensuremath{\,\mu{\rm s}}\xspace}    

\def\to                 {\ensuremath{\rightarrow}\xspace}

\def\pep2{PEP-II}

\def\gsim{{~\raise.15em\hbox{$>$}\kern-.85em
          \lower.35em\hbox{$\sim$}~}\xspace}
\def\lsim{{~\raise.15em\hbox{$<$}\kern-.85em
          \lower.35em\hbox{$\sim$}~}\xspace}

\def\P       {\ensuremath{P}\xspace}

\xspace

\newcommand{\jprlBase}       {Phys.\ Rev.\ Lett.\xspace}

\newcommand{\jprl}      [1]  {\jprlBase\ {\bf #1}}

\newcommand{\yf}        [1]  {{Yad.\ Fiz.\ {\bf #1}}}
\newcommand{\zp}        [1]  {\zpBase\ {\bf #1}}

\def\jetset74   {\mbox{\tt Jetset \hspace{-0.5em}7.\hspace{-0.2em}4}\xspace}

\def\beq{\begin{equation}}
\def\eeq{\end{equation}}

\def\beqa{\begin{eqnarray}}
\def\eeqa{\end{eqnarray}}

\def\bracket<#1|#2>{\setbox0=\vbox{\hbox{$#1$$#2$}}\left<#1\kern1pt \vrule  height\ht0\kern2pt #2\right>}

\def\rhop{{\ensuremath{\rho^+}}}
\def\rhom{{\ensuremath{\rho^-}}}

\def\rhopm{{\ensuremath{\rho^{\pm}}}}

\def\half{\ensuremath{{1\over 2}}}

\def\fs{\ensuremath{F}}    
\def\fsb{{\ensuremath{\kern 0.2em \overline{\kern -0.2em F}}}} 
\def\A#1{\ensuremath{A_{#1}}}
\def\r#1{R_{#1}}
\def\P{\ensuremath{P}}
\def\Pb{{\ensuremath{\kern 0.2em \overline{\kern -0.2em P}}}}
\def\Vb{{\ensuremath{\kern 0.2em \overline{\kern -0.05em V}}}}

\def\rhop{\ensuremath{\rho_+}}
\def\rhom{\ensuremath{\rho_-}}
\def\rhopm{\ensuremath{\rho_\pm}}

\def\rhob{\ensuremath{\bar\rho}}
\def\rhobp{\ensuremath{\rhob_+}}
\def\rhobm{\ensuremath{\rhob_-}}
\def\rhobpm{\ensuremath{\rhob_\pm}}

\def\tilderhobp{\ensuremath{\tilde{\rhob}_+}}
\def\tilderhobm{\ensuremath{\tilde{\rhob}_-}}
\def\tilderhobpm{\ensuremath{\tilde{\rhob}_\pm}}

\def\fD{\ensuremath{f_{\Dz}^{\fs}}}     
\def\fDb{\ensuremath{f_{\Dzb}^{\fs}}}   

\def\fDPrime{\ensuremath{f_{\Dz}^{\fs'}}} 
\def\fDbPrime{\ensuremath{f_{\Dzb}^{\fs'}}}

\def\fbD{\ensuremath{f_{\Dz}^{\fsb}}}   
\def\fbDb{\ensuremath{f_{\Dzb}^{\fsb}}} 

\def\rB{\ensuremath{r_B}}

\def\xp{\ensuremath{x_+}}
\def\xm{\ensuremath{x_-}}
\def\xpm{\ensuremath{x_{\pm}}}
\def\xz{\ensuremath{x_{\fs}}}
\def\xf#1{\ensuremath{x_{#1}}}

\def\yp{\ensuremath{y_+}}
\def\ym{\ensuremath{y_-}}
\def\ypm{\ensuremath{y_{\pm}}}
\def\yz{\ensuremath{y_{\fs}}}
\def\yf#1{\ensuremath{y_{#1}}}
\def\z{\ensuremath{z}}
\def\zp{\ensuremath{z_+}}
\def\zm{\ensuremath{z_-}}
\def\zpm{\ensuremath{z_{\pm}}}
\def\zz{\ensuremath{z_{\fs}}}
\def\zf#1{\ensuremath{z_{#1}}}
\def\Az{\ensuremath{A_0}}

\def\btou{\ensuremath{b\to u\cbar s}}
\def\btoc{\ensuremath{b\to c\ubar s}}
\def\ppp{\ensuremath{\pi^+\pi^-\pi^0}}

\def\btodk{\ensuremath{B^- \to D K^-}}

\def\bpmtodkpm{\ensuremath{B^\pm\to DK^\pm}}
 
\def\dtoppp{\ensuremath{D\to \ppp}}

\def\dztoppp{\ensuremath{\Dz\to \ppp}}

\def\BR#1{\ensuremath{{\cal B}(#1)}}
\def\Br{\ensuremath{{\Gamma}}}
\def\br{\ensuremath{\hat{\Gamma}}}
\def\Ntot{N}

\def\figurebox#1#2#3{%
    \def\arg{#3}%
    \ifx\arg\empty
    {\hfill\vbox{\hsize#2\hrule\hbox to #2{\vrule\hfill\vbox to #1{\hsize#2\vfill}\vrule}\hrule}\hfill}%
    \else
    {\hfill\epsfbox{#3}\hfill}%
    \fi}

\begin{document}

\title{\large \bf 
\boldmath
CP-violation parameters from decay rates of $B^\pm \to D K^\pm$,
$D\to $ multibody final states
}

\author{A.~Soffer}
\affiliation{Tel Aviv University, Tel Aviv, 69978, Israel}
\author{W.~Toki}
\affiliation{Colorado State University, Fort Collins, CO, 80523, USA}
\author{F.~Winklmeier}
\affiliation{CERN, CH-1211 Geneva 23, Switzerland}

\date{\today}

\begin{abstract}
We describe a method for measuring CP-violation parameters from which
the Cabibbo-Kobayashi-Maskawa angle $\gamma$ may be extracted. The
method makes use of the total decay rates in $B^\pm \to DK^\pm$ decays, where
the neutral $D$ meson decays to multibody final states. 
We analyze the
error of the method using experimental CP-violation analysis variables that 
enable straightforward sensitivity comparison with other methods
for extracting $\gamma$, and discuss the use of
$B$-factory and charm-factory data to obtain the relevant charm decay
information needed for this measurement.
Measurement sensitivities are estimated for the currently available
$B$-factory data sample, and $D$ decay modes for which use of this
method can make a significant contribution toward reducing the total
error on $\gamma$ are identified.

\end{abstract}
\pacs{13.25.Hw, 11.30.Er}

\maketitle
\vskip .3 cm

\section{Introduction}

An important part
of the program to study CP~violation is the measurement of
the angle $\gamma = \arg{\left(- V^{}_{ud} V_{ub}^\ast/ V^{}_{cd}
V_{cb}^\ast\right)}$ of the unitarity triangle related to the 
Cabibbo-Kobayashi-Maskawa (CKM) quark-mixing matrix~\cite{ref:km}.
Measurement of $\gamma$ performed with tree-level processes defines an
experimentally allowed region for the apex of the unitarity
triangle. This region should overlap with the region obtained from
$\Bz-\Bzb$ and $\Bs-\Bsb$ mixing, assuming there are no significant
new-physics contributions in the mixing amplitude.
With this assumption, current Tevatron measurements~\cite{ref:bsmix}
of the $\Bs-\Bsb$ mixing rate yield an indirect constraint on $\gamma$
that is much tighter than direct
measurements~\cite{ref:ckmfitter-utfit}. Therefore, precise direct
determination of $\gamma$ presents an opportunity to conduct an
accurate test of the Standard Model.

The decays $B\to DK$ can be used to measure $\gamma$ with
essentially no hadronic uncertainties, exploiting interference between
the \btou\ and \btoc\ amplitudes of the decays $B\to \Dzb K$ and $B \to \Dz
K$, respectively~\cite{Gronau:1991dp}.
Interference takes place when the $D$ meson\footnote{
We use the symbol $D$ to indicate any linear combination of
a $\Dz$ and a $\Dzb$ meson state.
} is observed in a final state $F$ that is accessible to both $\Dz$
and $\Dzb$ decays.
Such measurements can be conducted with quite a few $D$ and $B$ decay
modes, including those with excited charm and strange mesons, 
involving different methods for constructing and optimizing
CP-violation observables and measuring parameters related to $\gamma$.
In fact, there has been a healthy stream of new ideas in this area
since the basic method was first proposed in
1991~\cite{Gronau:1991dp}.  The different parameters of the various
measurements are then combined statistically, yielding confidence
intervals for $\gamma$~\cite{ref:ckmfitter-utfit}.
The statistical sensitivity provided by each mode and method is
generally poor, mainly due to the strong CKM suppression (and, for
most modes~\cite{ref:aps}, color suppression) in the \btou\
transition. This necessitates the exploitation of as many modes and methods
as possible, in order to achieve a small combined error on $\gamma$. 

The most accurate $\gamma$ measurement method to date 
determines $\gamma$ by analyzing
the $D$-decay event distribution in $\bpmtodkpm$ with multibody
$D$ decays~\cite{Giri:2003ty,bondar}.
This method was initially applied to the Cabibbo-favored decay $D\to
\KS \pi^+\pi^-$~\cite{Abe:2003cn,Aubert:2006am}, and 
the \babar\ Collaboration later used it with 
$\KS K^+K^-$~\cite{Aubert:2008bd} and 
the Cabibbo-suppressed decay \dtoppp~\cite{Aubert:2007ii}. 
A simulation study has also been conducted for the four-body mode
$D\to K^+K^-\pi^+\pi^-$~\cite{Rademacker:2006zx}.

As originally proposed~\cite{Giri:2003ty}, this method extracts the
angle $\gamma$ from measurements of $d\Br_\pm^{\fs}(\P)/d\P$, the
differential decay rates of $B^\pm \to D K^\pm$ at each phase-space 
point $\P$ of the multibody $D$-decay final state \fs.  However,
measurements done with the final states $\fs= \KS \pi^+\pi^-$ and
$\fs= \KS K^+K^-$ have only made use of the phase-space distributions,
given by the relative differential rates
\beq
{d\br_\pm^\fs(\P) \over d\P} \equiv { d\Br_\pm^\fs(\P) \over d\P} 
          \, {1 \over  \Br_\pm^\fs},
\eeq
where 
\beq
\Br_\pm^\fs \equiv \int  {d\Br_\pm^\fs(\P)\over d\P} d\P
\eeq
are the total decay rates.  Thus, these measurements were sensitive only
to the dependence of the rates on the point $\P$, not to their
integrated values $\Br_\pm^\fs$.
By contrast, the \babar\ measurement with $\fs=\ppp$ used both
$d\br_\pm^\fs(\P)/d\P$ and $\Br_\pm^\fs$. For that mode, the total
decay rate $\Br_\pm^\fs$ gave more precise information about the
CP-violation parameters than the phase-space distribution
$d\br_\pm^\fs(\P)/d\P$. While most measurements and sensitivity
estimates have focused on use of $d\br_\pm^\fs(\P)/d\P$ for learning
about $\gamma$, it is important to identify and study 
the decay modes for which the
total decay rate has competitive sensitivity to the CP-violation
parameters. This will help ensure that all useful modes are
utilized for measuring $\gamma$, while preventing much effort
from being wasted on data analysis of decay modes that are not promising.

The purpose of this paper is to provide the tools for estimating the
CP-parameter sensitivities of measurements of the absolute decay rates
$\Br_\pm^\fs$ for different $D$ decay modes. We demonstrate that
a good estimate of the sensitivities is provided by a single
mode-dependent parameter. 
The impact of each mode on the combined error of $\gamma$ depends on
values of strong phases and decay distributions that in many cases are
not well known yet. However, our general analysis of the
sensitivities, performed in terms of CP-violation parameters similar
to those used in the most accurate experimental analyses to date,
provides a good indication as to when using the integrated decay rates
is expected to improve the overall precision on $\gamma$.
Since the combined error on $\gamma$ depends on many measurements, its
full estimation is not within the scope of this paper and is not
attempted here. Rather, we compare the sensitivity of the
absolute-rate analysis to that of the current-best
phase-space-distribution analysis using comparable experimental CP-violation
variables.

We present the formalism for the decay rates in \btodk\ with multibody
$D$ decays in Section~\ref{sec:rates}.  Methods for measuring
important charm-decay quantities are discussed in
Section~\ref{sec:zz}.  The sensitivities with which the CP-violation
parameters are obtained from the total rates are calculated in
Section~\ref{sec:sensitivities}, then estimated for self-conjugate $D$
decay modes in Section~\ref{sec:rates-self} and for non-self-conjugate
modes in Section~\ref{sec:rates-non-self}. We provide numerical
estimates for several cases, in which enough information is available
for carrying out this calculation, indicating the promising and
not-so-promising final states for this type of analysis. Actual data
analysis of the type discussed here has been performed for only one of
the decay modes we study, $D\to \ppp$. For all other modes, the
estimates we provide are new.

\section{$\mathbf \btodk$ Decay Rates}
\label{sec:rates}
Consider the decay $B^\pm \to DK^\pm$, $D\to \fs(\P)$, where
$D$ is a superposition of the $\Dz$ and $\Dzb$ states, $\fs$ represents
the particles comprising a multibody final state accessible through
both $\Dz$ and $\Dzb$ decays, and $P$ is a specific point in the 
phase space of \fs.
We are also interested in events involving the
decay $D\to \fsb(\Pb)$, where $\fsb(\Pb)$ is the CP conjugate 
of $\fs(\P)$.
The $B$-meson decay amplitudes to final states with specific 
charm flavor are parameterized and denoted in this paper in the following way:
\beqa
A(B^-\to \Dz K^-) = A(B^+\to \Dzb K^+) &=& A_B,   \nonumber\\
A(B^-\to \Dzb K^-) &=& A_B \zm, \nonumber\\
A(B^+\to \Dz K^+) &=& A_B \zp, 
\label{eq:basic-b-decays}
\eeqa
where the complex numbers
\beq
\zpm \equiv \rB e^{i (\delta_B \pm \gamma)}
\label{eq:zpm-def}
\eeq
are the CP-violation parameters of interest, $\rB \sim 0.1$ is the
non-negative ratio between the magnitudes of the interfering \btou\
and \btoc\ amplitudes, and $\delta_B$ is the CP-even phase difference
between them.
The magnitude $|A_B|$ is measured~\cite{ref:b2dk} from the rate of the
process $B^-\to \Dz K^-$, $\Dz\to K^-\pi^+$, where contamination by the
interfering decay chain $B^-\to \Dzb K^-$, $\Dzb\to K^-\pi^+$ is
doubly Cabibbo-supressed as well as \rB-suppressed.

We define the magnitudes $\A{\fs}$ and $\r{\fs} \A{\fs}$
to be the square roots of the total \Dz\ decay rates into \fs\ and
\fsb,
\beqa
\A{\fs} &\equiv& \sqrt{\Gamma(\Dz\to \fs)} = \sqrt{\Gamma(\Dzb\to \fsb)}, 
   \nonumber\\[2mm]
\r{\fs} &\equiv& {1 \over \A{\fs}} \sqrt{\Gamma(\Dz\to \fsb)} 
      = {1 \over \A{\fs}} \sqrt{\Gamma(\Dzb\to \fs)}. 
\label{eq:A_F}
\eeqa
The ratio $\r{\fs}$ equals 1 for charge self-conjugate final states ($\fs =
\fsb$), but can in general have any non-negative value.
Eqs.~(\ref{eq:A_F}) ignore the possible impact of CP-violation in
$D$ decays. In addition, our use below of $\A{\fs}$ and $\r{\fs}$ will also
ignore the effect of $\Dz-\Dzb$ mixing. It has been
demonstrated~\cite{ref:dmix-gsz} that these effects can be neglected
for the purpose of measuring $\gamma$, as long as this is done
consistently for the $D$ mesons produced in the $B$ decay as well as for those
used to determine necessary $D$-decay quantities, discussed in 
Section~\ref{sec:zz}.
Alternatively, previously measured mixing and CP violation in
$D$ decays can be explicitly accounted for in the
formalism~\cite{ref:dmix-siso}. For the purpose of the current
discussion, it is sufficient to neglect these effects, as we do
throughout this paper.

We define the normalized amplitude distribution functions for the \P-dependent
charm meson decays,
\beqa
\fD(\P)    &\equiv& {A(\Dz\to\fs(\P)) \over \A{\fs}} , \nonumber\\[2mm]
\fDb(\P)   &\equiv& {A(\Dzb\to\fs(\P)) \over \A{\fs} \r{\fs}} , \nonumber\\[2mm]
\fbD(\Pb)  &\equiv& {A(\Dz\to\fsb(\Pb)) \over \A{\fs} \r{\fs}},
     \nonumber\\[2mm]
\fbDb(\Pb) &\equiv& {A(\Dzb\to\fsb(\Pb)) \over \A{\fs} }.
\label{eq:DAmps}
\eeqa
These functions satisfy the relations
\beq
\fbD(\Pb) = \fDb(\P), \ \ \ \ \  \fbDb(\Pb) = \fD(\P)
\label{eq:noCPinD}
\eeq
as a result of CP conservation in the charm meson decays, and 
are explicitly normalized, such that
\beq
\int \left|\fD(\P)\right|^2 d\P =
\int \left|\fDb(\P)\right|^2 d\P = 1.
\label{eq:func-norm}
\eeq

Accounting for the interference between the \btou\ and \btoc\
amplitudes in the $B$ meson decays, the amplitudes for the four full
decay chains are obtained from Eqs.~(\ref{eq:basic-b-decays}),
(\ref{eq:A_F}), and~(\ref{eq:DAmps}),
\beqa
A(B^- \to \fs(\P) K^-) &=& \Az \left(\fD(\P) + \r{\fs} \fDb(\P) z_- \right),
   \nonumber\\
A(B^+ \to \fs(\P) K^+) &=& \Az \left(\r{\fs} \fDb(\P) + \fD(\P) z_+ \right),
   \nonumber\\
A(B^- \to \fsb(\Pb) K^-) &=& \Az \left(\r{\fs} \fbD(\Pb) + \fbDb(\Pb) z_- 
                                                                   \right),
   \nonumber\\
A(B^+ \to \fsb(\Pb) K^+) &=& \Az \left(\fbDb(\Pb) + \r{\fs} \fbD(\Pb) z_+ 
                                                                   \right),
\nonumber\\
\label{eq:basic-amps}
\eeqa
where $\Az \equiv |A_B| \A{\fs}$. 
The observable \P-dependent $B$-decay rates are the
squares of these amplitudes, 
\beqa
{d\Br_\pm^{\fs}(\P)\over d\P} &=& \left|A(B^\pm \to \fs(\P) K^\pm)\right|^2, 
   \nonumber\\
{d\Br_\pm^{\fsb}(\Pb)\over d\Pb} &=&\left|A(B^\pm \to \fsb(\Pb) K^\pm)\right|^2.
\label{eq:basic-rates-P}
\eeqa

In the case $\fsb=\fs$, namely, when the $D$ decay final state is
self conjugate, only two of the four
equations~(\ref{eq:basic-rates-P}) are unique. These are the modes
that have
been studied experimentally so
far~\cite{Abe:2003cn,Aubert:2006am,Aubert:2008bd,Aubert:2007ii,abe:2008wya}.
As mentioned above, measurements of \zpm\
using $\fs=\KS\pi^+\pi^-$ and $\fs=\KS K^+K^-$ have been performed by analyzing
only the \P-dependence of the event distributions $d\br_\pm(\P)/d\P$,
disregarding the total decay rates $\Br_\pm^\fs$.
Since fitting $d\br_\pm(\P)/d\P$ in terms of \rB, $\gamma$, and $\delta_B$
leads to an average upward bias in \rB\ when \rB\ is of order its
experimental error,
Refs.~\cite{Abe:2003cn,Aubert:2006am,Aubert:2008bd,abe:2008wya} used
the CP-violation parameters
\beq 
\xpm \equiv \Re\{\z_\pm\}, \ \ \ \ \
\ypm \equiv \Im\{\z_\pm\},
\eeq
which are unbiased for this type of analysis.
After these parameters are measured in the analysis of
$d\br_\pm^\fs(\P)/d\P$, they are converted into (in general, non-Gaussian)
confidence regions in terms of the ``physical'' parameters \rB,
$\gamma$, and $\delta_B$.

Here, however, we wish to focus on and generalize the approach used
experimentally in Ref.~\cite{Aubert:2007ii} and
first studied theoretically in Ref.~\cite{Atwood:2003mj},
by examining the additional information
that can be extracted from the total decay rates $\Br^\fs_\pm$ and
$\Br^\fsb_\pm$. The expressions for these rates 
are obtained by taking the squared absolute value of
Eqs.~(\ref{eq:basic-amps}) and integrating over all phase-space
points, 
\beqa
\Br^\fs_- 
    &=& \Az^2 \left(1 + \r{\fs}^2 |\zm|^2 - 2\r{\fs} \Re\{\zz^*\zm\}\right),
   \nonumber\\
\Br^\fs_+  
    &=& \Az^2 \left(\r{\fs}^2 + |\zp|^2 - 2\r{\fs} \Re\{\zz\zp\}\right),
   \nonumber\\
\Br^\fsb_-  
    &=& \Az^2 \left(\r{\fs}^2 + |\zm|^2 - 2\r{\fs} \Re\{\zz\zm\}\right),
   \nonumber\\
\Br^\fsb_+  
    &=& \Az^2 \left(1 + \r{\fs}^2 |\zp|^2 - 2\r{\fs} \Re\{\zz^*\zp\}\right),
\label{eq:basic-rates}
\eeqa
where
\beqa
\zz &\equiv& -\int \fD(\P) \left(\fDb(\P)\right)^* d\P \nonumber\\
    &=&      -\int \fD(\P) \left(\fbD(\Pb)\right)^* d\P
\label{eq:zz}
\eeqa
is a measure of the interference between the $\Dz$ and $\Dzb$ decay
amplitudes into the final state \fs, averaged over the final-state
phase space. The absolute value and argument of \zz\ are,
respectively, the coherence parameter and average strong phase of
Ref.~\cite{Atwood:2003mj}.  For the purpose of this discussion, it
will be more useful to graphically think of
\zz\ as a coordinate-system offset parameter for \zpm.
Methods to measure \zz\ are outlined in
Section~\ref{sec:zz}. The
important point for now is that \zz\ can be measured significantly more
precisely than \zpm\ from high-statistics $D$ decay samples, namely,
\beq
\sigma_{\zz} \ll \sigma_{\zpm}.
\label{eq:sigma_zz<<}
\eeq

It is useful to represent \zpm\ in terms of the parameters
\beqa
\rhopm  &\equiv& \zpm - {1 \over \r{\fs}} \zz , \nonumber\\
\rhobpm &\equiv& \zpm - \r{\fs} \zz^*.
\label{eq:cyl-coord}
\eeqa
We follow Ref.~\cite{Aubert:2007ii} in referring to \rhopm\ and
\rhobpm\ as the polar-coordinate parameters. This designation is
motivated by the fact that measurement of the absolute decay rates
is directly related to the radii $|\rhopm|$\ and
$|\rhobpm|$, via the relations
\beqa
\Br^{\fs}_-  
    &=& \Az^2 \left(1 + \r{\fs}^2 |\rhom|^2 - |\zz|^2 \right),
   \nonumber\\
\Br^\fs_+  
    &=& \Az^2 \left(\r{\fs}^2 + |\rhobp|^2 - \r{\fs}^2 |\zz|^2 \right),
   \nonumber\\
\Br^\fsb_- 
    &=& \Az^2 \left(\r{\fs}^2 + |\rhobm|^2 - \r{\fs}^2 |\zz|^2 \right),
   \nonumber\\
\Br^\fsb_+ 
    &=& \Az^2 \left(1 + \r{\fs}^2 |\rhop|^2 - |\zz|^2 \right).
\label{eq:rates-rho}
\eeqa
\begin{figure}[!htbp]
\begin{center}
\includegraphics[width=0.4\textwidth]{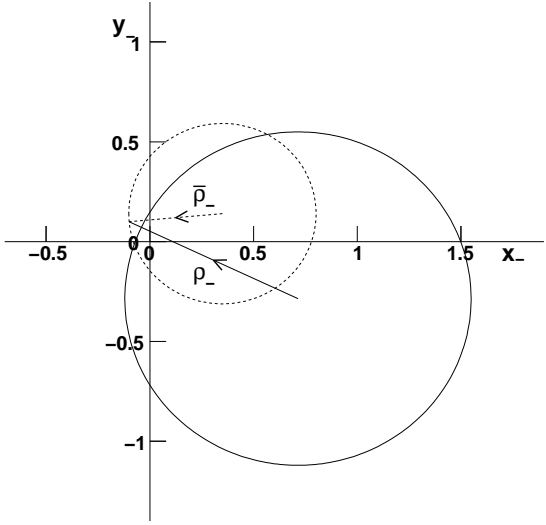}
\caption{The relationship 
  between the Cartesian and polar coordinates
  for $\zz = 0.5 - 0.2 i$, $\zm = -0.1 + 0.1 i$, $\r{\fs} = 0.7$. 
  The solid (dotted) arrow corresponds to the complex number
  \rhom\ (\rhobm) of Eq.~(\ref{eq:cyl-coord}).
  A measurement of $|\rhom|$ ($|\rhobm|$) implies that the true value of 
  \zm\ may be anywhere on the solid (dotted) circle in the $(\xm,\ym)$ plane. 
  The two crossing points of the circles are the possible solutions
  of \zm.
}
\label{fig:coords}
\end{center}
\end{figure}
Fig.~\ref{fig:coords} demonstrates the relationship between the
polar coordinates \rhom, \rhobm\ and the Cartesian coordinates
\xm, \ym\ for specific values of \zm\ and \zz.  
The absolute values $|\rhopm|$, $|\rhobpm|$
extracted from the total decay rates of Eqs.~(\ref{eq:rates-rho}) yield
two possible values for \zm\ and two for \zp, for a solution of
$\gamma$ with a four-fold ambiguity.  
In that sense, this is identical to measuring $\gamma$ with two,
two-body $D$ modes, as in the method of Ref.~\cite{ref:ads},
whose discrete ambiguities are further discussed in Ref.~\cite{Soffer:1999dz}.
In the case of multibody $D$ modes, analysis of the distribution of
events throughout the \fs\ phase space reduces the ambiguity to
two-fold, in addition to improving the total
precision~\cite{Giri:2003ty}. In effect, the event phase-space-distribution
analysis measures not only the absolute value but also the phase of
\rhopm\ and \rhobpm~\cite{Aubert:2007ii}. Combining the
phase-space-distribution analysis with the total-rates analysis yields the
most precise measurement of $\gamma$ for a given $D$ decay mode.

\section{Measuring \boldmath $\zz$}
\label{sec:zz}

A general approach for measuring the components of \zz\ from decay
rates of the $\psi(3770)$ into neutral-$D$ final states has been
developed in Ref.~\cite{Atwood:2003mj}. Consider the case in which one
of the $\psi(3770)$ daughters decays into $\fs(\P)$ and the other decays into
$\fs'(\P')$, where the phase-space points $\P$ and $\P'$ do not have
to be related.  Due to the quantum numbers $J^{PC} = 1^{--}$ of the
$\psi(3770)$, its two-$D$ decay wave function is antisymmetric under
exchange of the daughters, and hence must be $\left(\Dz\Dzb -
\Dzb\Dz\right)/\sqrt{2}$. The normalized event density in the
$\P\P'$ phase space is 
\beqa
&& {2 \over B_{\Dz\Dzb} \A{\fs}^2 \A{\fs'}^2}\,
     {d\Gamma_{\fs, \fs'}(\P, \P') \over d\P d\P'} 
    \nonumber\\ 
 &=&    \left| \fD(\P) \r{\fs'}{\fDbPrime}(\P') 
           - \r{\fs} \fDb(\P) {\fDPrime}(\P') \right|^2,
\label{eq:gen-density-3770}
\eeqa
where $B_{\Dz\Dzb}$ is the $\psi(3770)\to \Dz\Dzb$ branching fraction.
Integrating this expression over phase space yields the normalized rate,
\beqa
\tilde \Gamma_{\fs, \fs'} &\equiv&
   {2 \over B_{\Dz\Dzb} \A{\fs}^2 \A{\fs'}^2}\,
  \Gamma_{\fs, \fs'} 
    \nonumber\\ 
 &=& \r{\fs}^2 + \r{\fs'}^2
     - 2 \r{\fs} \r{\fs'} \Re\left\{\zz  \zf{\fs'}^* \right\}  \nonumber\\
 &=& \r{\fs}^2 + \r{\fs'}^2
     - 2 \r{\fs} \r{\fs'} \left(\xz \xf{\fs'} + \yz \yf{\fs'} \right),
\label{eq:gen-rate-3770}
\eeqa
where Eq.~(\ref{eq:zz}) was used, and we have separated \zz\ and
\zf{\fs'} into their real and imaginary parts,
\beqa
\zz &\equiv& \xz + i \yz, \nonumber\\
\zf{\fs'} &\equiv& \xf{\fs'} + i \yf{\fs'}.
\eeqa

By measuring the decay rate of Eq.~(\ref{eq:gen-rate-3770}) for
different final states $\fs'$, one obtains all the information about
\zz. We begin with $\fs' = \fs$, for which Eq.~(\ref{eq:gen-rate-3770}) becomes
\beqa
\tilde \Gamma_{\fs, \fs} &=& 2 \r{\fs}^2 \left(1 - |\zz|^2 \right)
   \nonumber\\
 &=& 2 \r{\fs}^2 \left(1 - \xz^2 - \yz^2  \right).
\label{eq:rate-same}
\eeqa

Next, we take $\fs'$ to be a CP-even or CP-odd state, namely
\beq 
\fs' = \Dz_\pm = {1 \over \sqrt{2}} \left(\Dz \pm \Dzb\right).
\label{eq:cpes}
\eeq
The inverse relations of Eq.~(\ref{eq:cpes}) yield
\beqa
\zf{\Dz_\pm} &=& \pm 1, \nonumber\\
\r{\Dz_\pm} &=& 1,
\eeqa
where the normalization condition Eq.~(\ref{eq:func-norm})
has been taken into account. Then
Eq.~(\ref{eq:gen-rate-3770}) becomes
\beq
\tilde \Gamma_{\fs, D_\pm} = 2 \left(1 \mp \xz \right).
\label{eq:rate-cpes}
\eeq

Eqs.~(\ref{eq:rate-same}) and~(\ref{eq:rate-cpes}) are sufficient for
obtaining \xz\ and \yz, the latter with a sign ambiguity. To
resolve this ambiguity, we now take $\fs'$ to be the 2-body state
$K^-\pi^+$.  
Eq.~(\ref{eq:gen-rate-3770}) then gives
\beqa
\tilde\Gamma_{\fs ,K^-\pi^+} &=& 
  \r{\fs}^2 + \r{K^-\pi^+}^2 \nonumber\\
     &-& 2 \r{\fs} \r{K^-\pi^+} \left(\xz \xf{K^-\pi^+} + \yz \yf{K^-\pi^+} 
                       \right).
\eeqa
We have yet to determine $\xf{K^-\pi^+}$ and $\yf{K^-\pi^+}$.
These are obtained from the rates 
\beqa
\tilde\Gamma_{K^-\pi^+ , K^-\pi^+} &=& 
   2 \r{K^-\pi^+}^2 \left(1 - \xf{K^-\pi^+}^2 - \yf{K^-\pi^+}^2  \right), 
      \nonumber\\
\tilde\Gamma_{K^-\pi^+  , D_\pm} &=& 2 \left(1 \mp \xf{K^-\pi^+} \right),
\eeqa
as in Eqs.~(\ref{eq:rate-same}) and~(\ref{eq:rate-cpes}), respectively.
Since $K^-\pi^+$ is a two-body state, $f_{\Dz}^{K^-\pi^+}$ and 
$f_{\Dzb}^{K^-\pi^+}$ are numbers
rather than functions. 
It then follows from Eq.~(\ref{eq:func-norm}), that $\zf{K^-\pi^+}$ 
has unit magnitude, and the constraint
\beq
\xf{K^-\pi^+}^2 + \yf{K^-\pi^+}^2 = 1,
\label{eq:kpi-unit-z}
\eeq
resolves the ambiguity in $\yf{K^-\pi^+}$. 
Thus, it is possible to measure the real and imaginary parts of $\zz$ with no
ambiguities.

Several studies~\cite{Soffer:1998un,Giri:2003ty,Bondar:model-indep}
have shown that when obtaining $D$-decay parameters from $\psi(3770)$
decays, the expected error on the CP-violation parameters due to 
the finite $\psi(3770)$ statistics is relatively small, given 
current CESR-c and $B$-factory integrated luminosities. 
A detailed
simulation study~\cite{Bondar:model-indep} has shown that in the
phase-space-distribution analysis with $\fs=\KS \pi^+\pi^-$, the error
on $\gamma$ due to the finite $\psi(3770)$ statistics is about four
times smaller than the error due to the finite \btodk\ statistics in
the currently available, $\sim 1~{\rm ab}^{-1}$ $B$-factory data
sample. Measurements 
performed by CLEO-c with 818~pb$^{-1}$ of 
$\epem\to\psi(3770)$ data have yielded an estimated 
$\gamma$ error of $1^\circ-2^\circ$ due to the measurement of the 
$D\to K_S\pi^+\pi^-$ decay 
parameters~\cite{cleoc:model-indep}. 
CLEO-c has also measured $\zz$ for the modes $K^-\pi^+\pi^0$
and $K^-\pi^+\pi^-\pi^+$,
obtaining the preliminary results~\cite{ref:libby}
$|\zf{K^-\pi^+\pi^0}| = 0.79 \pm 0.08$, 
$\arg\{\zf{K^-\pi^+\pi^0}\} = (197 {+28\atop -27})^\circ$,
$|\zf{K^-\pi^+\pi^+\pi^-}| = 0.24 {+0.21 \atop -0.17}$, 
$\arg\{\zf{K^-\pi^+\pi^0}\} = (161 {+85 \atop -48})^\circ$.
We note that while these errors are large, 
their impact on the errors of $|\rhobpm|$, which are
the relevant CP-violation parameters for small-$\r{\fs}$ modes
(see Section~\ref{sec:rates-non-self}) is suppressed by 
$\r{\fs}$, as seen from Eqs.~(\ref{eq:cyl-coord}).

Furthermore, the newly launched BEPC-II charm factory, with a design
luminosity almost twenty times that of CESR-c, will be able to
supply the charm data needed to match the large $B$ samples
that will be collected at LHCb and possibly at a proposed $e^+e^-$ ``super
$B$ factory''.
We conclude that the error on $\zpm$ in a decay-rate analysis of the
type presented in this paper will be dominated by the experimental
error on $|\rhopm|$ and not by knowledge of \zz.

\subsection{Self-Conjugate Modes}
\label{sec:self-conj}

So far, all the multibody $D$-decay modes studied experimentally
within the context of $B^\pm\to DK^\pm$ have been charge
self conjugate, i.e., $\fs=\fsb$.  From Eq.~(\ref{eq:A_F}), one
sees that self-conjugate modes satisfy $\r{\fs}=1$.
In addition, Eq.~(\ref{eq:noCPinD}), together with the condition
$\fs=\fsb$, implies 
\beq
\fD(\Pb) = \fDb(\P).
\label{eq:special-equality-for-self-conj}
\eeq
As a result, such states satisfy
\beq
\yz = 0, 
\label{eq:y0}
\eeq
as we demonstrate by dividing the phase space of $\fs$ into two
equal-volume regions $V$ and $\Vb$, such that every point $\P \in V$
is related to a point $\Pb \in \Vb$ by the CP transformation. 
For example, in a three-body decay of the type 
$D\to a^+ a^- b^0$, the division is
along the line $(p_{a^+} + p_{b^0})^2 = (p_{a^-} + p_{b^0})^2$,
where $p_j$ is the four-momentum of particle $j$. Such a
division can be performed for any multibody final state that is
self conjugate, regardless of its particle multiplicity. Then
\beqa
\yz &=& \Im\left\{ \int_V \fD(\P)\, \left(\fDb(\P)\right)^* d\P \right\}
      \nonumber\\
   &+& \Im\left\{\int_{\Vb} \fD(\Pb) \, \left(\fDb(\Pb)\right)^* d\Pb  \right\}.
\nonumber\\
\label{eq:y0=0_1}
\eeqa
Using Eq.~(\ref{eq:special-equality-for-self-conj}), the second integral
in Eq.~(\ref{eq:y0=0_1}) can be written as
\beq
\Im\left[\int_V \fDb(\P) \, \left(\fD(\P) \right)^* d\P \right].
\label{eq:2nd-term}
\eeq
The integrand in Eq.~(\ref{eq:2nd-term}) is the complex conjugate of
the integrand of the first term in Eq.~(\ref{eq:y0=0_1}). Therefore,
their imaginary parts cancel in the sum, yielding
$\yz=0$.

In the $\gamma$-related measurements performed so far with $B^\pm\to
DK^\pm$ and $D\to\fs$ decays into a multibody, self-conjugate state, a
particular model was assumed for the functional form of $\fD(\P)$. The
parameters of the model were obtained by fitting the phase-space
distribution of $D^0\to\fs$ decays, where the flavor of the $D^0$ was
tagged by its production in the decay $D^{*+}\to\Dz\pi^+$.  In this
case, Eq.~(\ref{eq:special-equality-for-self-conj}) guarantees that
\zz\ can be fully determined by inserting the model $\fD(\P)$ into
Eq.~(\ref{eq:zz}), as done in Ref.~\cite{Aubert:2007ii}.  
The same cannot be done for modes that are not
self conjugate, where one must resort to the use of $\psi(3770)$
decays.

\section{Experimental Sensitivities}
\label{sec:sensitivities}

Due to the linear relationship~(\ref{eq:rates-rho}) between the
experimentally observable rates and $|\rhopm|^2$,
$|\rhobpm|^2$, these squared radii are the unbiased CP-violation parameters of
choice for the rates analysis, given that decay rates can almost
always be obtained from reasonably unbiased estimators. If the errors
on $|\rhopm|^2$ and $|\rhobpm|^2$ are significantly smaller than the values of
these parameters, then their roots $|\rhopm|$ and $|\rhobpm|$ are also
unbiased parameters.

In terms of the errors on the rates, the errors on $|\rhopm|$ and
$|\rhobpm|$ are
\beqa
\sigma_{|\rhom|} &=& {\sigma_{\Br_-^\fs} \over 2 |\rhom| \Az^2 \r{\fs}^2},
    \nonumber\\[1mm]
\sigma_{|\rhobp|} &=& {\sigma_{\Br_+^\fs} \over 2 |\rhobp| \Az^2},
    \nonumber\\[1mm]
\sigma_{|\rhobm|} &=& {\sigma_{\Br_-^\fsb} \over 2 |\rhobm| \Az^2},
    \nonumber\\[1mm]
\sigma_{|\rhop|} &=& {\sigma_{\Br_+^\fsb} \over 2 |\rhop| \Az^2 \r{\fs}^2},
\label{eq:rho-errors}
\eeqa
where we have used Eq.~(\ref{eq:sigma_zz<<}) to neglect the error on $|\zz|$.
As a result of Eqs.~(\ref{eq:sigma_zz<<}) and~(\ref{eq:cyl-coord}),
the errors on $|\rhopm|$ and $|\rhobpm|$ are similar in magnitude to
the errors on \xpm\ and \ypm.  Therefore, studying 
Eqs.~(\ref{eq:rho-errors}) provides a simple means to compare the
$\gamma$ sensitivity of a rates analysis using any final state \fs\ to
the sensitivity of the current-best measurement, namely, that of \xpm\
and \ypm\ from the phase-space-distribution analysis of $\fs=\KS\pi^+\pi^-$.
In what follows, we make quantitative estimates of the errors on 
$|\rhopm|$ and $|\rhobpm|$.

\subsection{Self-conjugate modes}
\label{sec:rates-self}

As a result of Eq.~(\ref{eq:y0}),
Eq.~(\ref{eq:cyl-coord}) simplifies to
\beq
\rhobpm = \rhopm = \zpm - \xz
\label{eq:simplified-rhopm-def}
\eeq
for self-conjugate modes.
Therefore, the two circles of Fig.~\ref{fig:coords} collapse onto each
other, and the rates measurement of the radii $|\rhopm|$ is no longer
sufficient for fully determining \zpm. This is hardly a problem, for
two reasons.  First, the phases of $\rhopm$ may be determined from the
event-distribution analysis, as was done in Ref.~\cite{Aubert:2007ii},
yielding a measurement of \zpm\ whose precision is enhanced due to the
use of all available experimental information. Second, as
stated in the introduction, precise knowledge of $\gamma$ can in any
case be obtained only by combining many measurements of parameters
related to $\gamma$. Therefore, measurement of $|\rhopm|$ helps reduce
the overall error on $\gamma$, even if it is not sufficient for
extracting $\gamma$ without information obtained from other
$\gamma$-related measurements.

Since \xz\ is well known, it is useful to estimate the errors on
$|\rhopm|$ for relevant $D$ decay modes, as they will correspond
closely to the errors on \zpm. Since we are dealing with the case
$\r{\fs}=1$, Eqs.~(\ref{eq:rho-errors}) become
\beq
\sigma_{|\rhopm|} = {\sigma_{\Br_\pm} \over 2 |\rhopm| \Az^2}
 = {1 + |\rhopm|^2 - \xz^2 \over 2 |\rhopm|} \cdot
   {\sigma_{\Br_\pm} \over \Br_\pm},
\label{eq:sigmarho}
\eeq
where we have used $\Br_\pm \equiv \Br_\pm^\fs = \Br_\pm^\fsb$, and
the second equality of Eq.~(\ref{eq:sigmarho}), obtained from
Eq.~(\ref{eq:rates-rho}), conveniently relates the $|\rhopm|$ errors
to the relative errors on the signal branching fractions.
We rely on previous ``reference'' experimental studies of the relevant
decay modes to obtain these relative errors for any hypothetical value of
$|\rhopm|$.  Suppose that in a reference measurement performed with
$B$-factory data of integrated luminosity $\tilde L$, one observed $\tilde
\Ntot_\pm$ signal $B^\pm\to\fs K^\pm$ events, from which the
rates $\tilde \Br^{\fs}_\pm$ were determined and the 
CP-violation parameter values $|\tilde\rhopm|^2$ were calculated. Let
$\tilde\Ntot \equiv \tilde\Ntot_+  +  \tilde\Ntot_-$.
Then the numbers of signal events that would be observed in an
experimentally identical, hypothetical measurement of luminosity $L$
given hypothetical values $|\rhopm|^2$ for the CP-violation
parameters, are
\beq
N_\pm = \tilde\Ntot \tilde r_\pm {L \over \tilde L}, 
\label{eq:future-br}
\eeq
where
\beq
\tilde r_\pm \equiv {\Br^{\fs}_\pm
                     \over \tilde \Br^{\fs}_- + \tilde \Br^{\fs}_+}
  = {1 + |\rhopm|^2 - \xz^2 \over 
                              2 + |\tilde\rhom|^2 + |\tilde\rhop|^2 - 2 \xz^2}
\label{eq:tilder}
\eeq
is the ratio between the value of the rate $\Br^{\fs}_\pm$ given the
hypothetical parameter values $\rhopm$ and the sum of the rates
$\tilde\Br^{\fs}_- + \tilde\Br^{\fs}_+$ measured in the reference
measurement.  The second equality in Eq.~(\ref{eq:tilder}) arises from
Eqs.~(\ref{eq:rates-rho}).

We assume that the error on the number of events $\tilde\Ntot$ in the
reference measurement can be written as the sum in quadrature of a
Poisson signal part and a background part, namely,
\beq
\sigma_{\tilde\Ntot}^2 = \tilde\Ntot + \sigma_{\tilde\Ntot, bgd}^2.
\label{eq:bgd-error-ref}
\eeq
Using this relation and the published reference-measurement quantities 
$\tilde\Ntot$ and $\sigma_{\tilde\Ntot}$, we obtain the
background contribution to the error, which we assume to be CP symmetric.
Then the errors on the numbers of events in the hypothetical
measurement, in which $N_\pm$ will be
observed, are
\beq
\sigma_{N_\pm} =
  \sqrt{{1\over 2}\left(\sigma_{\tilde\Ntot}^2 - \tilde\Ntot\right)  
         + \tilde\Ntot \tilde r_\pm}\, \sqrt{L \over \tilde L},
\label{eq:future-error}
\eeq
where the statistical assumption leading to
Eq.~(\ref{eq:bgd-error-ref}) was again used.  From
Eqs.~(\ref{eq:future-br}) and~(\ref{eq:future-error}), we obtain the
relative branching-fraction errors for the hypothetical measurement,
\beq
{\sigma_{\Br_\pm} \over \Br_\pm} = {\sigma_{N_\pm} \over N_\pm} = 
 {
  \sqrt{{1\over 2}\left(\sigma_{\tilde\Ntot}^2 - \tilde\Ntot\right)  
         + \tilde\Ntot \tilde r_\pm}
   \over
  \tilde\Ntot \tilde r_\pm
} \sqrt{\tilde L \over L}.
\label{eq:ratio-final}
\eeq

In Table~\ref{tab:x0} we report \xz, $|\rhopm|$, and
$\sigma_{|\rhopm|}$ for several three-body $D$ decay modes, assuming a
data sample of $10^9$ $e^+e^-\to B\bar B$ events, similar to the
currently available $B$-factory sample.
We obtain the values of \xz\ from Eq.~(\ref{eq:zz}), using the
Dalitz-plot distributions $\fD(\P)$, whose parameterizations are
reported in Refs.~\cite{Aubert:2007ii}, \cite{Abe:2003cn},
\cite{Aubert:2008bd}, and~\cite{Aubert:2007dc} for the $D$-decay final
states \ppp, $\KS \pi^+\pi^-$, $\KS K^+ K^-$, and $K^+ K^- \pi^0$,
respectively. We also obtain $\tilde\Ntot$, $\sigma_{\tilde\Ntot}$,
and $\tilde L$ from these references, except for $K^+ K^- \pi^0$,
where we estimate $\tilde\Ntot$ and $\sigma_{\tilde\Ntot}$ from
their values in \ppp~\cite{Aubert:2007ii}, taking into account the
ratio of branching fractions $\BR{\Dz\to K^+ K^- \pi^0} /
\BR{\dztoppp}$~\cite{ref:dppp} and an assessment that the background
yield in $K^+ K^- \pi^0$ will be 20\% of that in \ppp.
Since extraction of $|\tilde\rhopm|^2$ from the total rates has been
reported only for the $\fs=\ppp$ mode~\cite{Aubert:2007ii}, we take
$|\tilde\rhopm| = |\rhopm|$ when evaluating $\tilde r_\pm$ for all
other modes, for lack of a better value.
We take the hypothetical CP-violation parameter values $|\rhopm|$
from the averages of the values of $\xpm$ and $\ypm$ reported in
Refs.~\cite{Aubert:2008bd} and~\cite{abe:2008wya}, 
\beqa
\xm &=&  0.097 \pm 0.034, \nonumber\\
\ym &=&  0.054 \pm 0.058, \nonumber\\
\xp &=& -0.087 \pm 0.031, \nonumber\\
\yp &=& -0.038 \pm 0.042.
\label{eq:currentxy}
\eeqa
The errors of Eq.~(\ref{eq:currentxy}) reflect the sensitivity of a
measurement conducted with $1.04\times 10^9$ $e^+e^-\to B\bar B$
events, comparable to the value used to produce Table~\ref{tab:x0}.

\begin{table}
\begin{center}
\caption{\label{tab:x0}Values of the inputs to Eq.~(\ref{eq:sigmarho})
and the expected errors $\sigma_{|\rhopm|}$ for different $D$ decay modes
and a $B$-factory data sample of $10^9$ $\epem\to\BB$ events,
calculated with the CP-violation parameters of Eq.~(\ref{eq:currentxy}).}
\begin{tabular}{l|r|r|r|r|r||r|r}
\hline\hline
Mode &  $\xz$  
     &  $\sigma_{\Br_-} \over \Br_-$   &  $\sigma_{\Br_+} \over \Br_+$ 
     & $|\rhom|$   &   $|\rhop|$  
     &  $\sigma_{|\rhom|}$   &   $\sigma_{|\rhop|}$  \\
\hline
$\pi^+\pi^-\pi^0$  & $ 0.85$ & $0.13$  & $0.10$  & $0.75$ & $0.94$ & $0.07$ & $0.06$ \\
$\KS \pi^+\pi^-$   & $-0.02$ & $0.05$ & $0.05$ & $0.13$ & $0.08$ & $0.19$ & $0.32$ \\
$\KS K^+K^-$       & $-0.31$ & $0.09$ & $0.10$ & $0.41$ & $0.23$ & $0.12$ & $0.20$ \\
$K^+K^-\pi^0$      & $ 0.20$ & $0.11$  & $0.11$  & $0.12$ & $0.29$ & $0.48$ & $0.19$ \\
\hline\hline
\end{tabular}
\end{center}
\end{table}

One can see from Eq.~(\ref{eq:sigmarho}), that the error
$\sigma_{|\rhopm|}$ is small when $|\rhopm|$ is large. Large $|\rhopm|$
requires \xz\ to be large, by virtue of Eq.~(\ref{eq:simplified-rhopm-def})
and the smallness of $|\zpm|$, demonstrated in Eq.(\ref{eq:currentxy}).
We note that some insight into the value of \xz\ for a particular mode
can be obtained by studying the distribution of events in the
$\Dz$-decay Dalitz plot, since generally, high level of apparent
symmetry under the exchange of the two charged particles leads to a
high value of \xz.

It is evident from Table~\ref{tab:x0} that of the three-body modes
studied here, only $\xf{\dtoppp}$ is large enough for
Eq.~(\ref{eq:sigmarho}) to yield $|\rhopm|$ errors that are
competitive with the errors of Eq.~(\ref{eq:currentxy}).  In
particular, the high-statistics, low-background mode $\KS\pi^+\pi^-$
ends up having large $|\rhopm|$ errors due to the very 
small value of $\xf{\KS\pi^+\pi^-}$.  On
the other hand, we expect that the methods for suppression of the
significant background in \ppp, which were first developed in
Ref.~\cite{Aubert:2005hi}, will improve in upcoming analyses. That
should reduce $\sigma_{|\rhopm|}$ for this mode below the simple
extrapolation shown in Table~\ref{tab:x0}.

\subsection{Non-self-conjugate modes}
\label{sec:rates-non-self}

We proceed to estimate
the errors on \rhopm\ and \rhobpm\ in $D$-decay final states that are
not self-conjugate, i.e., $\fs \ne \fsb$.
As in the procedure leading up to
Eq.~(\ref{eq:sigmarho}), we replace $\Az^2$ in
Eqs.~(\ref{eq:rho-errors}) using
Eq.~(\ref{eq:rates-rho}):
\beqa
\sigma_{|\rhom|} &=& {\nu_{\rhom} \over 2|\rhom|\r{\fs}^2}
\cdot {\sigma_{\Br_-^\fs} \over \Br_-^\fs},
\nonumber\\[2mm]
\sigma_{|\rhobp|} &=& {\bar\nu_{\rhobp} \over 2|\rhobp|}
  \cdot {\sigma_{\Br_+^\fs} \over \Br_+^\fs},
\nonumber\\[2mm]
\sigma_{|\rhobm|} &=& {\bar\nu_{\rhobm} \over 2|\rhobm|}
  \cdot {\sigma_{\Br_-^\fsb} \over \Br_-^\fsb},
\nonumber\\[2mm]
\sigma_{|\rhop|} &=& {\nu_{\rhop} \over 2|\rhop|\r{\fs}^2}
  \cdot {\sigma_{\Br_+^\fsb} \over \Br_+^\fsb},
\label{eq:sigma-rho-nonself}
\eeqa
where 
\beqa
\nu_{\rhopm} &\equiv& 1 + \r{\fs}^2 |\rhopm|^2 - |\zz|^2, \nonumber\\
\bar\nu_{\rhobpm} &\equiv& \r{\fs}^2 + |\rhobpm|^2 - \r{\fs}^2 |\zz|^2.
\label{eq:nu}
\eeqa
As in Eq.~(\ref{eq:ratio-final}), the relative errors in
Eqs.~(\ref{eq:sigma-rho-nonself}) are obtained from the number of
signal events $\tilde\Ntot^\fs$, $\tilde\Ntot^\fsb$ and their errors,
observed in existing reference measurements,
\beq
{\sigma_{\Br_\pm^\fs} \over \Br_\pm^\fs} = 
    {\sqrt{\half\left(\sigma_{\tilde\Ntot^\fs}^2 - \tilde\Ntot^\fs\right)
	   + \tilde\Ntot^\fs \tilde r_\pm^\fs}
     \over \tilde\Ntot^\fs \tilde r_\pm^\fs} \sqrt{\tilde L\over L},
\label{eq:relerr-ns}
\eeq
with an analogous expression for \fsb,
where by analogy with Eq.~(\ref{eq:tilder}),
\beqa
\tilde r_-^\fs &\equiv&  
   {\nu_{\rhom} \over \nu_{\tilde\rhom} + \bar\nu_{\tilderhobp}}, 
     \nonumber\\
\tilde r_+^\fs &\equiv&  
   {\bar\nu_{\rhobp} \over \nu_{\tilde\rhom} + \bar\nu_{\tilderhobp}}, 
     \nonumber\\
\tilde r_-^\fsb &\equiv& 
   {\bar\nu_{\rhobm} \over \nu_{\tilde\rhop} + \bar\nu_{\tilderhobm}}, 
     \nonumber\\
\tilde r_+^\fsb &\equiv& 
   {\nu_{\rhop} \over \nu_{\tilde\rhop} + \bar\nu_{\tilderhobm}}.
\label{eq:tilder-nonself}
\eeqa
As in Section~\ref{sec:rates-self}, the symbols $\tilde\rhopm$,
$\tilderhobpm$ in Eq.~(\ref{eq:tilder-nonself}) refer to the
CP-violation parameters extracted from the reference measurements
$\tilde\Ntot^\fs$ and $\tilde\Ntot^\fsb$.  If the total rates were
not used to extract CP-violation parameters, one can naively take 
$\tilde\rhopm$ and $\tilderhobpm$ from Eq.~(\ref{eq:currentxy})
for the purpose of performing this error estimate.

Let us consider this error estimate in the case of the
non-self-conjugate, three-body final state $\fs = \KS K^- \pi^+$.
With as little as 5\% of their currently available data sample, the
\babar\ collaboration has performed a preliminary analysis of this
mode's Dalitz-plot amplitude-distribution functions $\fD(\P)$ and
$\fbD(\Pb)$~\cite{Aubert:2002yc}, from which we compute 
$|\zz| = 0.47$.
The ratio $\r{{\KS} K^-\pi^+} = 0.68$ is easily extracted from the
results reported in Ref.~\cite{Aubert:2002yc}. With $\r{{\KS} K^-\pi^+}$
being different from 1 yet of order 1, this mode is in a class of
Cabibbo-suppressed decays expected to exhibit large interference
between the \btou\ and \btoc\
decays~\cite{Grossman:2002aq}. Unfortunately, as we show below, the
combination of a small branching fraction and a medium-sized $|\zz|$
render $\KS K^- \pi^+$ unattractive for extracting $\gamma$ via the
total-rate method.

In addition to $\r{{\KS} K^-\pi^+}$ and $|\zf{{\KS} K^-\pi^+}|$, 
calculation of all four
errors of Eq.~(\ref{eq:sigma-rho-nonself}) also requires knowledge of
$\arg\{\zf{\KS K^-\pi^+}\}$, which has not been measured. However, a rough
estimate of the CP-parameter errors shows them to be comparable to
those of the $\KS K^+K^-$ mode, due to the following two observations.
First, the combined branching fraction
$\BR{\Dz\to \KS K^-\pi^+} + \BR{\Dz\to \KS K^+\pi^-}$
is approximately 85\% of $\BR{\KS K^+K^-}$. One therefore expects the
relative error on $\Ntot^{\KS K\pi}$ to be somewhat larger than that
on $\Ntot^{\KS K^+K^-}$. Experimental details, such as kaon vs. pion
multiplicities and combinatoric background under the larger $\KS K\pi$
Dalitz plot, slightly increase our expectation for the ratio between
the relative errors on $\Ntot^{\KS K\pi}$ and $\Ntot^{\KS K^+K^-}$.
The second observation is that $|\zf{\KS K^-\pi^+}|$ is about 50\%
larger than $\xf{\KS K^+K^-}$.  Combining these two competing
effects, we conclude that the errors on components of the CP-violation
parameters obtained from $\KS K\pi$ and $\KS K^+K^-$ should be of
similar magnitudes. As seen in Table~\ref{tab:x0}, this implies
error values that are too large to be of practical interest.

We note that Eqs.~(\ref{eq:sigma-rho-nonself}) also hold for
Cabibbo-allowed final states involving a single charged kaon, such as
$\fs=K^-\pi^+\pi^0$, for which $\r{K^-\pi^+\pi^0} \approx
0.05$~\cite{ref:pdg06} (where we have ignored the effect of $\Dz-\Dzb$
mixing~\cite{ref:dmix-gsz}).  Eqs.~(\ref{eq:rates-rho}) show that in
this case, the sensitivity of $\Br^{\fs}_-$ and $\Br^{\fsb}_+$ to the
CP-violation parameters is suppressed by $\r{K^-\pi^+\pi^0}^2$, making
these rates useful for obtaining $A_B$, as mentioned in
Section~\ref{sec:rates} for the $\Dz\to K^-\pi^+$ decay.
However, the absolute rates $\Br^{\fs}_+$ and $\Br^{\fsb}_-$ do
provide a good measurement of $|\rhobpm|$. 
Searching for these decays in
a data sample of $226\times 10^6$ $e^+e^-\to B\bar B$ events,
\babar~\cite{Aubert:2007nu} has put an upper limit on the ratio
\beq
R_{ADS} \equiv {\Br^{\fsb}_- + \Br^{\fs}_+ \over  \Br^{\fs}_- + \Br^{\fsb}_+}
 = {\bar\nu_{\rhobm} + \bar\nu_{\rhobp} \over \nu_{\rhom} + \nu_{\rhop}},
\label{eq:rads}
\eeq
for which the central value obtained was $\tilde R_{ADS} = 0.013 {+
0.010 \atop - 0.004}$.  
The rates that appear in the
numerator of Eq.~(\ref{eq:rads}), to which we refer as
the ADS rates~\cite{ref:ads}, are suppressed by factors of second order
in the small parameters $\rB$, $\r{K^-\pi^+\pi^0}$ relative to the rates in the
denominator.  The error on $R_{ADS}$ is dominated by the statistical
errors on the ADS rates.
To properly account for this when calculating the relative errors on
the ADS rates, we evaluate Eq.~(\ref{eq:relerr-ns}) with
\beq
\tilde r_\pm^{K\pi\pi^0} = 
  {\bar\nu_{\rhobpm} \over \bar\nu_{\tilderhobm} + \bar\nu_{\tilderhobp}}
\eeq
instead of the expressions in Eq.(\ref{eq:tilder-nonself}), and take
$\tilde N^{K\pi\pi^0}$ to be the number of ADS events detected in
Ref.~\cite{Aubert:2007nu}, namely, $19 \pm 10$, where the 10-event error is
obtained from the naive average of the positive and negative errors on
$\tilde R_{ADS}$.

\begin{figure}[!htbp]
\begin{center}
\includegraphics[width=0.45\textwidth]{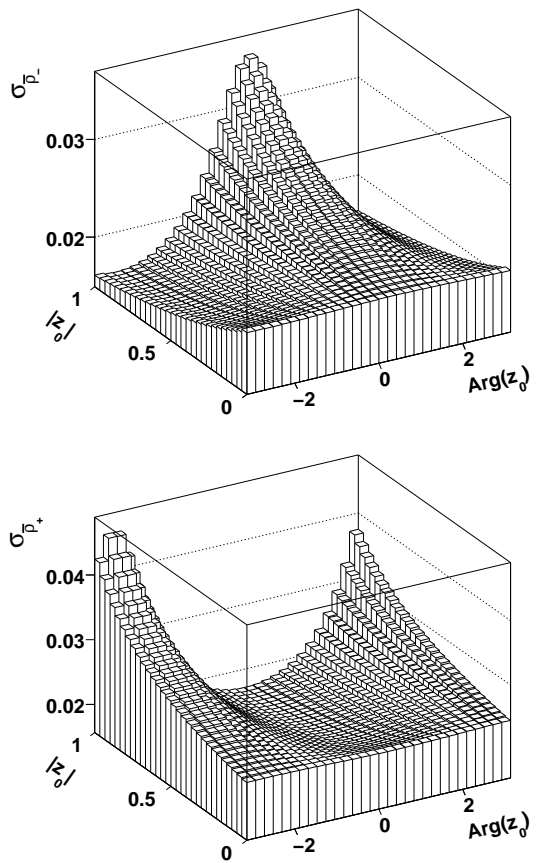}
\caption{The errors on $|\rhobm|$ (top) and $|\rhobp|$ (bottom) as functions
   of the absolute value and phase of \zz\ for $\fs=K^-\pi^+\pi^0$,
   calculated for $10^9$ $\epem\to\BB$ events with the CP-violation
   parameters of Eq.~(\ref{eq:currentxy}).}
\label{fig:kpipi0} 
\end{center}
\end{figure}

The resulting errors on $|\rhobpm|$ are shown in
Fig.~\ref{fig:kpipi0}, calculated with
Eqs.~(\ref{eq:sigma-rho-nonself}) for different values of \zz. As in
the case of Table~\ref{tab:x0}, we have assumed a data sample of
$10^9$ $\epem\to\BB$ events and the CP-violation parameter values of
Eq.~(\ref{eq:currentxy}).  The errors reach values as low as $0.016$
and as high as $0.035$ ($0.045$) for $|\rhobm|$ ($|\rhobp|$).  We see
that at least one of the errors is smaller than about 0.025 for any
value of \zz.  For the CLEO-c central values of
$\zf{K^-\pi^+\pi^0}$~\cite{ref:libby}, we find 
$\sigma_{|\rhobpm|} \approx 0.02$.

These results suggest that one can expect measurement of
the CP-violation parameters with $\fs=K^-\pi^+\pi^0$ to yield errors
that are very competitive with the current-best measurement,
Eq.~(\ref{eq:currentxy}), once the luminosity is high enough for
observation of the ADS decays. 
%

\section{Discussion}
\label{sec:disc}

Of the self-conjuage final states studied quantitatively here,
the errors obtained from \ppp\ are the smallest, due to the large
value of $\xz=\Re\{\zz\}=0.85$ in this mode.  The errors are expected
to decrease beyond the estimate shown in Table~\ref{tab:x0}, as
background suppression improves in subsequent analyses of this mode.
By contrast, the final state $\KS\pi^+\pi^-$, which thanks in part to
its large branching fraction and high purity has yielded the most
precise phase-space-distribution measurements of $\gamma$ to date,
has a very small \xz, rendering its absolute decay rates
poor measures of the CP-violation parameters.

Our calculations show that measuring $|\rhobpm|$ with the final state
$K^-\pi^+\pi^0$ can yield very small errors, smaller than or of
similar magnitude to the errors from the phase-space-distribution
analysis of $\KS\pi^+\pi^-$.  We note that similar 
precision may be obtained with the two- and four-body final states $K^-\pi^+$
and $K^-\pi^+\pi^-\pi^+$,
whose study is outside the scope of this paper.

The results presented here cover the major three-body
$D$-decay final states with known and significant branching fractions. It
is possible that the absolute decay rates into some of the
higher-multiplicity states will also turn out to yield competitive errors
on $\gamma$.
Among the Cabibbo-suppressed modes, this includes the final state
$K^+K^-\pi^+\pi^-$, whose phase-space-distribution analysis has been
studied in simulation~\cite{Rademacker:2006zx}, and $2\pi^+2\pi^-$.
The Cabibbo-favored mode $\Dz\to \KS\pi^+\pi^-\pi^0$ has a large
branching fraction, $(5.3 \pm 0.6)\%$~\cite{ref:pdg06}, and may
therefore be attractive for both phase-space-distribution and
absolute-decay-rate analyses.  Since almost half the rate is due to
the resonant contribution $K^{*-}(892) \rho^+$, the phase-space
distribution is highly asymmetric under exchange of the two charged
pions. Therefore, it is unlikely that \xz\ is large for this
mode. Nonetheless, given the large branching fraction, even $\xz$ as
small as $0.1$ could make this mode attractive for studying $\gamma$.

\section{Conclusions}
\label{sec:conc}

We have studied the use of the absolute $B^\pm \to DK^\pm$ decay rates,
where the $D$ decays to a multibody final state, for obtaining
information with which to improve the overall knowledge of the CKM
unitarity-triangle phase $\gamma$. This information is complementary
to that obtained from other $\gamma$-related measurements, including 
analysis of the $\Dz-\Dzb$ interference pattern seen
in the phase-space distributions of the $D$ decay products.
We have developed a 
formalism for estimating the error on the CP-violating parameters
$|\rhopm|$ and $|\rhobpm|$. The parameter that most
strongly affects the errors is \zz\ of Eq.~(\ref{eq:zz}). We have 
evaluated \zz\ for three-body $D$ final states for which the necessary input
information is available, and have estimated the errors on the CP-violation
parameters for these self-conjugate modes and for the modes
$\KS K^-\pi^+$ and $K^-\pi^+\pi^0$.

\begin{acknowledgments}
This research was supported in part
by grant number 2006219 from the United States-Israel Binational
Science Foundation (BSF), Jerusalem, Israel. 
The authors thank Werner Sun, Jure Zupan, and Jim Libby 
for useful suggestions.

\end{acknowledgments}


\begin{thebibliography}{99}

\bibitem{ref:km} N.~Cabibbo, \jprl {\bf 10},
        531 (1963);
  M.~Ko\-ba\-yashi and T.~Maskawa, 
        Prog. Theoret. Phys.  {\bf 49}, 652 (1973).


\bibitem{ref:bsmix} The CDF - Run II Collaboration (A.~Abulencia {\it et al.}),
  Phys.\ Rev.\ Lett.\  {\bf 97}, 062003 (2006)
  [arXiv:hep-ex/0606027].


\bibitem{ref:ckmfitter-utfit} The CKMfitter Group (J.~Charles {\it et al.}), 
  Eur.\ Phys.\ J.\  C {\bf 41}, 1 (2005)
  [arXiv:hep-ph/0406184],
  updated results: http://ckmfitter.in2p3.fr;
The UTfit Collaboration (M.~Bona {\it et al.}),
  JHEP {\bf 0610}, 081 (2006)
  [arXiv:hep-ph/0606167],
  updated results: http://www.utfit.org/.


\bibitem{Gronau:1991dp}
M.~Gronau and D.~Wyler, 
Phys.\ Lett.\ B {\bf 265},  172 (1991).


\bibitem{ref:aps} But see 
R.~Aleksan, T.~C.~Petersen and A.~Soffer,
  Phys.\ Rev.\  D {\bf 67}, 096002 (2003)
  [arXiv:hep-ph/0209194];
R.~Aleksan and T.~C.~Petersen,
   {\it In the Proceedings of 2nd Workshop on the CKM Unitarity Triangle, 
   Durham, England, 5-9 Apr 2003, pp WG414}
  [arXiv:hep-ph/0307371].



\bibitem{Giri:2003ty}
A.~Giri,  Y.~Grossman,  A.~Soffer and J.~Zupan, 
Phys.\ Rev.\ D {\bf 68},  054018 (2003)
[arXiv:hep-ph/0303187].


\bibitem{bondar}
A.~Bondar, Proceedings of BINP Special Analysis Meeting on Dalitz 
Analysis, 24-26 Sep. 2002, unpublished.



\bibitem{Abe:2003cn}
The Belle Collaboration, A.~Poluektov {\it et al.},
  Phys.\ Rev.\ D {\bf 73}, 112009 (2006)
  [arXiv:hep-ex/0604054].


\bibitem{Aubert:2006am}
The \babar\ Collaboration, B.~Aubert {\it et al.},
  Phys.\ Rev.\ Lett.\  {\bf 95}, 121802 (2005)
  [arXiv:hep-ex/0504039].




\bibitem{Aubert:2008bd} The \babar\ Collaboration (B.~Aubert {\it et al.}),
 Phys.\ Rev.\  D {\bf 78}, 034023 (2008)
 [arXiv:0804.2089 [hep-ex]].




\bibitem{Aubert:2007ii}
The \babar\ Collaboration, B.~Aubert {\it et al.},
  Phys.\ Rev.\ Lett.\  {\bf 99}, 251801 (2007)
  [arXiv:hep-ex/0703037].


\bibitem{Rademacker:2006zx}
  J.~Rademacker and G.~Wilkinson,
  Phys.\ Lett.\  B {\bf 647}, 400 (2007)
  [arXiv:hep-ph/0611272].


\bibitem{ref:b2dk}  The CLEO Collaboration (M.~Athanas {\it et al.}),
  Phys.\ Rev.\ Lett.\  {\bf 80}, 5493 (1998)
  [arXiv:hep-ex/9802023];
The \babar\ Collaboration (B.~Aubert {\it et al.}),
 Phys.\ Rev.\ Lett.\  {\bf 92}, 202002 (2004)
 [arXiv:hep-ex/0311032].



\bibitem{ref:dmix-gsz}
  Y.~Grossman, A.~Soffer and J.~Zupan,
  Phys.\ Rev.\  D {\bf 72}, 031501(R) (2005)
  [arXiv:hep-ph/0505270].

\bibitem{ref:dmix-siso}
J.~P.~Silva and A.~Soffer,
  Phys.\ Rev.\  D {\bf 61}, 112001 (2000)
  [arXiv:hep-ph/9912242].


\bibitem{abe:2008wya}
The Belle Collaboration,  K.~Abe {\it et al.},
  arXiv:0803.3375 [hep-ex].




\bibitem{Atwood:2003mj}
  D.~Atwood and A.~Soni,
  Phys.\ Rev.\  D {\bf 68}, 033003 (2003)
  [arXiv:hep-ph/0304085].



\bibitem{ref:ads} D.~Atwood, I.~Dunietz and A.~Soni,
  Phys.\ Rev.\ Lett.\  {\bf 78}, 3257 (1997)
  [arXiv:hep-ph/9612433].


\bibitem{Soffer:1999dz}
  A.~Soffer,
  Phys.\ Rev.\  D {\bf 60}, 054032 (1999)
  [arXiv:hep-ph/9902313].


\bibitem{Soffer:1998un}
  A.~Soffer,
  arXiv:hep-ex/9801018.



\bibitem{ref:cleo-4170}
  D.~M.~Asner and W.~M.~Sun,
  Phys.\ Rev.\  D {\bf 73}, 034024 (2006)
  [Erratum-ibid.\  D {\bf 77}, 019902 (2008)]
  [arXiv:hep-ph/0507238].


\bibitem{Bondar:model-indep}
  A.~Bondar and A.~Poluektov,
  Eur.\ Phys.\ J.\  C {\bf 47}, 347 (2006)
  [arXiv:hep-ph/0510246];
A.~Bondar and A.~Poluektov,
  arXiv:0801.0840 [hep-ex].

\bibitem{cleoc:model-indep} J.~Rademacker, 
   {\it In the Proceedings of 5th International 
    Workshop on the CKM Unitarity Triangle, 
    Rome, Italy, 9-13 Sept, 2008}.


\bibitem{ref:libby} J.~Libby
   {\it In the Proceedings of 5th International 
    Workshop on the CKM Unitarity Triangle, 
    Rome, Italy, 9-13 Sept, 2008}.





\bibitem{Aubert:2007dc}
The \babar\ Collaboration,  B.~Aubert {\it et al.},
  Phys.\ Rev.\  D {\bf 76}, 011102 (2007)
  [arXiv:0704.3593 [hep-ex]].


\bibitem{ref:dppp}
The \babar\ Collaboration, B.~Aubert  {\it et al.}, 
  Phys.\ Rev.\ D {\bf 74}, 091102 (2006)
  [arXiv:hep-ex/0608009].




\bibitem{Aubert:2002yc}
The \babar\ Collaboration, B.~Aubert {\it et al.},
 arXiv:hep-ex/0207089.



\bibitem{Grossman:2002aq}
  Y.~Grossman, Z.~Ligeti and A.~Soffer,
  Phys.\ Rev.\ D {\bf 67}, 071301(R) (2003)
  [arXiv:hep-ph/0210433].


\bibitem{ref:pdg06}
Particle Data Group, Y.-M.~Yao {\it et al.},  
J.\ Phys.\ G {\bf 33}, 1 (2006).



\bibitem{Aubert:2007nu}
The \babar\ Collaboration,  B.~Aubert {\it et al.},
  Phys.\ Rev.\  D {\bf 76}, 111101 (2007)
  [arXiv:0708.0182 [hep-ex]].


\bibitem{Aubert:2005hi}
The \babar\ Collaboration, B.~Aubert {\it et al.},
  Phys.\ Rev.\ D {\bf 72}, 071102 (2005)
  [arXiv:hep-ex/0505084].








\end{thebibliography}
\end{document}